\documentstyle[preprint,aps,epsf]{revtex}

\begin{document}

\title{
{\large\hfill NBI preprint Nr.\ 96--41}\\
In-medium Production of Kaons at the Mean-Field Level}

\author{
J\"urgen Schaffner$^a$, Jakob Bondorf$^a$, 
and Igor N. Mishustin$^{a,b}$}
\address{
$^a$The Niels Bohr Institute,
Blegdamsvej 17,
DK-2100 Copenhagen}
\address{
$^b$The Kurchatov Institute Russian Research Center,
Moscow 123182} 

\maketitle

\begin{abstract}
The in-medium mass and energy of kaons and antikaons 
are studied within the Relativistic Mean 
Field approach and compared with predictions from chiral models
by taking care of kaon-nucleon scattering data.
Implications for the subthreshold production of kaons and antikaons 
in heavy-ion collisions are discussed. 
We find only small corrections due to in-medium effects
on the mean-field level for the relevant production processes for 
kaons. The production of kaons is even less favourable 
at high density due to repulsive vector interactions.
We conclude that one has to go beyond mean-field approaches
and take fluctuations and secondary production processes
into account to explain the recently measured enhancement of kaon 
production at subthreshold energies.
The situation is different for antikaons where in-medium effects
strongly enhances their production rates.
We also see strong in-medium modifications of the annihilation
processes of antikaons and $\Lambda$'s which might be visible
in flow measurements. 
At high density, we predict that 
the threshold energy for antikaon and $\Lambda$
production and annihilation become equal leading to similar numbers
of antikaons and $\Lambda$'s 
in the dense zone of a relativistic heavy ion collision.
\end{abstract}

\section{Introduction}

In-medium properties of hadrons have received 
considerable attention recently, both experimentally and theoretically
by studying relativistic heavy-ion collisions. 
Charged kaons (K$^+$) seem to be a quite promising tool for probing
the dense interior of the collision zone as their mean free path
is long enough to escape without further interactions.
Kaplan and Nelson proposed first that a kaon condensed phase may be formed
in the dense matter created in heavy ion collisions \cite{Nel87}.
Further studies within the Nambu--Jona-Lasinio model \cite{Lutz92}, 
chiral perturbation theory \cite{Brown94} and an one-boson exchange model
\cite{Sch94b} showed that the kaon (K$^+$) sees a repulsive
potential in the medium and will not condense. 

On the other side, 
the antikaon (K$^-$) feels a strong attraction 
which is confirmed by recent calculations taking into account the
contribution coming from the $\Lambda(1405)$ resonance just below 
threshold \cite{Koch94,Waas96}.
It was then predicted by chiral perturbation theory that a antikaon condensed
phase will form in the dense interior of a neutron star 
\cite{Brown92} consistent with scattering data
\cite{Brown94,Lee94} and Kaonic atoms \cite{Lee95}.
This approach has been criticised in \cite{Sch94b,Maru94}
as the scalar density is set equal to the baryon density and higher
order terms in density are neglected.
The appearance of hyperons \cite{Prak} shifts also the onset of a condensed
phase to higher density. 
As shown in  \cite{Sch96}
a strong nonlinear dependence on density
and the implementation of 
hyperon-hyperon interactions even prevents an antikaon condensed
phase inside a neutron star.

Here we will continue our work for neutron stars \cite{Sch96} and
apply it for the situation in heavy ion collisions at threshold.
Subthreshold production rates of K$^+$ in heavy-ion
reactions were recently measured at GSI \cite{GSI}.
Earlier work showed that the in-medium modifications  
of kaons and antikaons might be measurable
in heavy ion collisions at threshold.
For example, it was shown that the kaons are sensitive to the equation of
state (EOS) \cite{Aich85,Hart94,Li95}. A softer EOS produces more
kaons than a hard one. On the other hand it will also depend
on the parametrisation used for
the cross section \cite{Rand80,Zwer88} but not on N-body collisions
\cite{Batko92} and not on the high-momentum tail of the nucleons
\cite{Sib95}.
The influence of rescattering and formation of resonance 
($\Delta$) matter was studied in the QMD model \cite{Hart94},
and the RBUU model \cite{Maru94b} and it was demonstrated that they
are essential to explain the data.
In-medium modifications
of the effective energy of the kaon were studied in \cite{Fang94}
using again the RBUU model. The results are essential similar
to the ones obtained without medium modifications \cite{Maru94b},
because the in-medium kaon mass
used is quite close to the respective vacuum mass.
But there exist other observables which might be better suited
for extracting in-medium effects.
The flow of kaons might be a promising tool for
measuring the kaon potential in dense matter \cite{kaonflow}.
And more pronounced in-medium effects are expected for the case of
K$^-$ \cite{Li94}. Indeed, enhanced production rates for K$^-$
have been seen at GSI recently \cite{GSI2}.

In this paper we want to examine the possible influence of a dense
nuclear environment on the properties of kaons and antikaons.  We show
that the in-medium effects on the mean-field level can not explain the
measured enhanced production rates of kaons in contrary to the
conclusion drawn in ref.\ \cite{Li94}.  We discuss two different
approaches: first an one-boson exchange model and second a chiral
approach where the parameters are fixed by s-wave scattering lengths
and the low density theorem.  In-medium effects for $\Lambda$'s are
also taken into account by linking them to hypernuclear data.  We show
that the phase space in the medium does not change considerably for
the processes NN$\to$N$\Lambda$K and secondary processes as
$\pi$N$\to\Lambda$K and N$\Delta\to$N$\Lambda$K due to cancelation
effects.  On the contrary, effects nonlinear in density even cause an
enhanced repulsion at highly dense matter for these processes. Hence,
subthreshold production of kaons seems {\em not} to probe the
potentials of the very dense region of a heavy-ion collision.  On the
other hand, in-medium effects are essential for explaining the
enhanced production of antikaons.  We show that the process
NN$\to$NNK$\bar{\rm K}$ is enhanced in the medium while the
annihilation process $\bar K$N$\to\Lambda\pi$ is reduced.
We also find that 
the annihilation process $\Lambda$N$\to$NN$\bar K$ is essentially
enhanced in dense matter and might be equally important as
antikaon annihilation.
This behaviour may lead to equal numbers of antikaons and $\Lambda$'s
in the dense zone of a relativistic heavy ion collision.

The paper is organized as follows: 
first we introduce the Relativistic Mean Field (RMF) model and extend it
to include $\Lambda$'s. In the second section we discuss two different
interaction schemes for the kaons with nuclear matter, one based on 
an one-boson exchange model 
and the other on chiral perturbation theory (ChPT). 
The parameters are fixed to the s-wave KN scattering lengths.
Results for the in-medium effects on kaon and antikaon 
production are presented in the third section. 
The last section is devoted to conclusions and an outlook.

\section{The RMF model}

The RMF model has been proven to give a good description of
nuclear matter in bulk and of the properties of finite nuclei 
\cite{Ser86,Rei89}.
We start from the Lagrangian
\begin{eqnarray}
{\cal L} &=&
\overline{\Psi}_N(i \gamma^\mu\partial_\mu - m_N)\Psi_N
+ \frac{1}{2}\partial^\mu \sigma \partial_\mu \sigma
-  U(\sigma) \cr\cr
&&
-\frac{1}{4}G^{\mu\nu}G_{\mu\nu}
+ \frac{1}{2}m_\omega^2 V^\mu V_\mu
-\frac{1}{4}\vec{B}^{\mu\nu}\vec{B}_{\mu\nu}
+ \frac{1}{2}m_\rho^2 \vec{R}^\mu \vec{R}_\mu \cr\cr
&&
- g_{\sigma N}\overline{\Psi}_N\Psi_N\sigma
- g_{\omega B}\overline{\Psi}_N\gamma^\mu\Psi_N V_\mu
- g_{\rho N}\overline{\Psi}_N\gamma^\mu
                        \vec{\tau}\Psi_N\vec{R}_\mu
\end{eqnarray}
where the nucleons interact via an attractive scalar ($\sigma$)
and a repulsive vector ($V^\mu $) meson field.
The term $U(\sigma)$ stands for the scalar selfinteraction
\begin{equation}
U(\sigma) = \frac{1}{2}m_\sigma^2 \sigma^2
+ \frac{b}{3}\sigma^3 + \frac{c}{4}\sigma^4
\label{eq:bognl}
\end{equation}
introduced by Boguta and Bodmer \cite{Bog77} to get a correct
compressibility of nuclear matter (for another stabilized functional form
see \cite{Rei88}).
The parameters of this Lagrangian can be fixed to 
bulk properties \cite{Bog77} or to the
properties of finite nuclei \cite{Rufa88,Shar93}.
A general discussion about the scalar 
selfinteraction terms can be found
in \cite{Bod89}.
Bodmer proposed an additional selfinteraction
term for the vector field \cite{Bod91}
\begin{equation}
{\cal L}_{V^4} = \frac{1}{4} d (V_\mu V^\mu)^2
\end{equation}
which leads to a soft equation of state at high densities
in agreement with Dirac-Br\"uckner calculations \cite{Bod91,Gmu91}.
Fits to the properties of nuclei with this new term  are quite
successful \cite{Toki94}.
In the following we take mostly the parameter sets NL-Z \cite{Rufa88}
which is the commonly used parameter set NL1 with a better
zero-point energy correction
and the recent set TM1 \cite{Toki94} with vector selfinteraction terms.
The former one gives a rather stiff equation of state while the latter
one a rather soft one.

The implementation of hyperons proceeds as
\begin{equation}
{\cal L}_\Lambda =
\overline{\Psi}_\Lambda (i \gamma^\mu\partial_\mu - m_\Lambda)\Psi_\Lambda
- g_{\sigma \Lambda}\overline{\Psi}_\Lambda \Psi_\Lambda \sigma
- g_{\omega \Lambda}\overline{\Psi}_\Lambda \gamma^\mu\Psi_\Lambda  V_\mu
\end{equation}
and the two new coupling constants can be fixed to hypernuclear data 
\cite{Bro77}.
The main feature of hypernuclei is that
the depth of the $\Lambda$-potential is about
\begin{equation}
U_\Lambda^{(N)} =
g_{\sigma\Lambda}\sigma^{\rm eq.} + g_{\omega\Lambda} V_0^{\rm eq.}
\approx - 30 \mbox{ MeV }
\label{eq:lambdapot}
\end{equation}
in saturated nuclear matter 
which already fixes one coupling constant of the $\Lambda$,
say $g_{\sigma\Lambda}$ \cite{Glen91,Sch92}.
The vector coupling constant $g_{\omega\Lambda}$ 
is then given by
\begin{equation}
g_{\omega \Lambda} = \frac{2}{3} g_{\omega N} 
\end{equation}
when using SU(6)-symmetry (the quark model, see e.g.\ \cite{Dov82}).
The SU(6)-symmetry also secures that the spin-orbit force is negligible small
as there is no experimental evidence for a spin-orbit splitting for
hypernuclear levels.
Noble showed first \cite{Nob80} that 
the contribution of the vector terms to the
spin-orbit term nearly cancel each other
when taking into account the tensor force
and SU(6)-symmetry. The tensor force vanishes 
in bulk matter on the mean field level 
as it is proportional to the gradient of the fields. 
Therefore it is not considered here.

\begin{figure}
\epsfbox{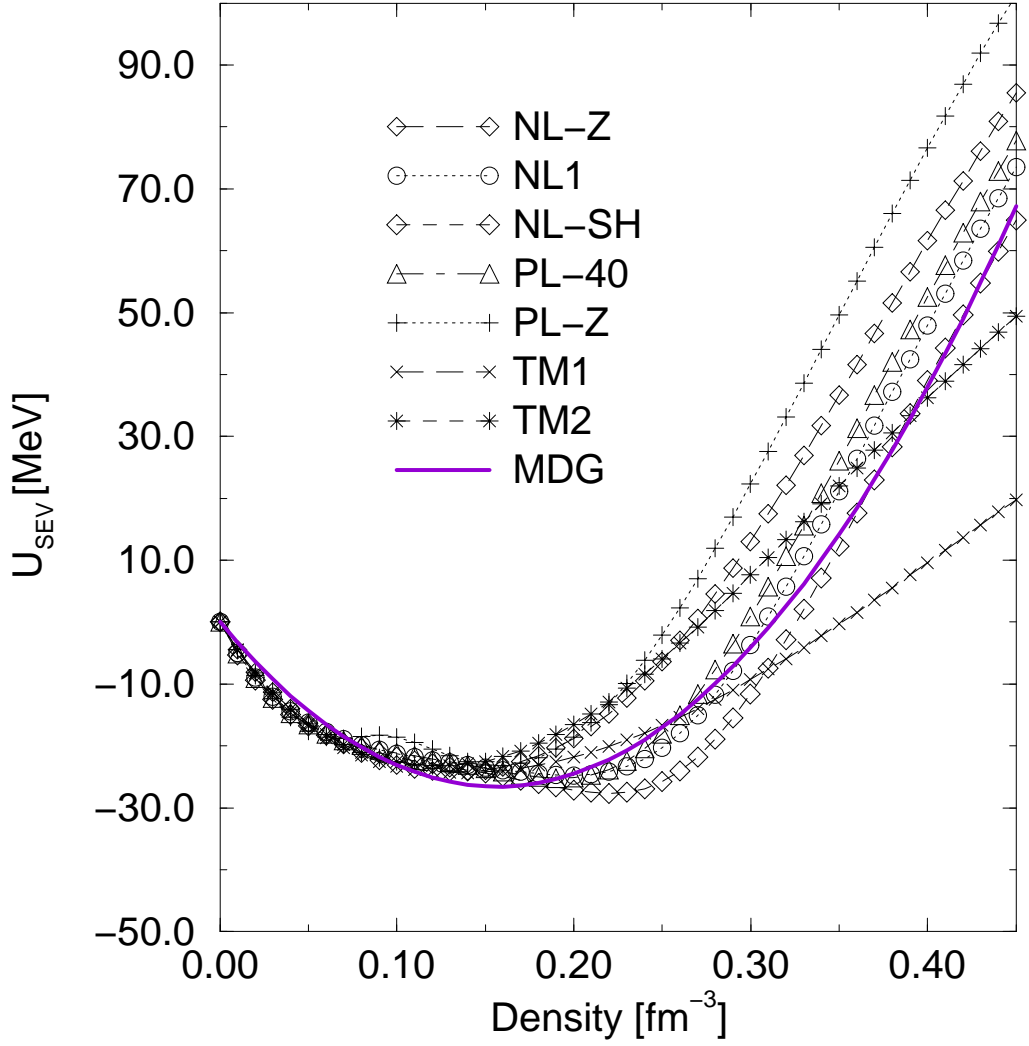}
\caption{
The Schr\"odinger equivalent potential of the $\Lambda$ 
for several parameter sets of the RMF model as a function of density.
The curve labelled MDG is the non-relativistic potential fit to
hypernuclear data of Dover, Millener and Gal \protect\cite{MDG88}. 
}
\label{fig1}
\end{figure}

The in-medium energy of nucleons and hyperons is then given by
\begin{eqnarray}
E_N (p) &=& \sqrt{(m_N + g_{\sigma N}\sigma)^2 + p^2} 
+ g_{\omega N} V_0 + g_{\rho N} \tau_0 R_{0,0} 
\label{eq:nucmass}\\
E_\Lambda (p) &=& \sqrt{(m_\Lambda + g_{\sigma \Lambda}\sigma)^2 + p^2} 
+ g_{\omega \Lambda} V_0 \quad .
\end{eqnarray}
It is important to note that the parameters here are connected to properties 
at normal nuclear matter density. 
The in-medium effects for
nucleons and $\Lambda$'s at this point are known and should be taken
into account when studying the influences and the signals 
of a dense nuclear environment.  
As pointed out in \cite{MDG88} three-body forces are also important to
explain hypernuclear data. As these forces are repulsive, the
hyperon potential shows a nonlinear behaviour with density and changes
sign at higher density. Figure~\ref{fig1} shows the Schr\"odinger equivalent
potential defined as
\begin{equation}
U_{\rm SEV} = 
g_{\sigma \Lambda}\sigma + g_{\omega \Lambda} V_0
+ \frac{1}{2m_\Lambda} \left(
(g_{\sigma \Lambda}\sigma)^2 - (g_{\omega \Lambda} V_0)^2 \right)
\end{equation}
for different parameter sets 
in comparison with the findings of the non-relativistic approach
\cite{MDG88}. 
The overall behaviour is quite similar despite of the EOS used. 
The nonlinear behaviour of
the scalar field with density seems to simulate the repulsive
three-body force of the nonrelativistic approach.
It also demonstrate that is crucial to make a difference between
scalar and vector (baryon) density.
This turning of the hyperon potential will be quite important
for our discussion of the kaon production in the medium.

\section{Kaon interactions}

The case for the kaon is quite distinct from that of the $\Lambda$.
There does not exist any kaon-nuclear states similar to hypernuclei
as the KN-interaction is known to be repulsive. 
Taking the (real) isospin averaged KN-scattering length 
$\bar a_{KN}=(3a_0^{I=1}+a_0^{I=0})/4=-0.255$ fm \cite{Bar94}
and using the low density theorem
one gets a repulsive optical potential depth at normal nuclear 
density of about
\begin{equation}
U^{\rm KN}_{\rm opt} = -\frac{2\pi}{m_K} \left(1+\frac{m_K}{m_N}\right)
\bar a_{KN} \rho_N \approx + 29 \mbox{ MeV } \rho_N / \rho_0
\label{eq:kaonpot}
\end{equation}
compatible with kaon (K$^+$) scattering on nuclear targets \cite{Dov82}.
Here we have taken the groundstate density to be $\rho_0 = 0.15$ fm$^{-3}$. 
The repulsive interaction is the reason 
why kaons have a long mean-free path in nuclear matter.
Note that the potential depth is just opposite 
to the one of the $\Lambda$ which signals
a significant cancellation of attractive and repulsive terms in the medium.
On the other hand, a recent experiment measured an enhanced cross section
for K$^+$ scattering on nuclear targets \cite{Weiss94} incompatible with
multiple scattering arguments. This is the so called K$^+$-puzzle
which is still unresolved.
The isospin dependent potential in nuclear matter can be estimated from
the isospin scattering length 
$a_{\rm iso}=(a_0^{I=1}-a_0^{I=0})/4=-0.055$ fm
and the low density theorem
\begin{equation}
U^{\rm KN}_{\rm iso} = -\frac{2\pi}{m_K} \left(1+\frac{m_K}{m_N}\right)
\bar a_{\rm iso} \rho_{\rm iso} \approx + 6 \mbox{ MeV } 
\rho_{\rm iso} / \rho_0
\label{eq:isoopt}
\end{equation}
where $\rho_{\rm iso}$ is the isospin density of the system.
For led one can estimate 
$\rho_{\rm iso} \approx (2Z-A)/A \rho_N \approx -0.21\rho_N$,
which gives about 1 MeV correction at normal nuclear density.

For antikaons the annihilation processes
\begin{equation} 
\bar K + {\rm N}  \to {\rm Y} + \pi \qquad ({\rm Y}=\Lambda , \Sigma )
\end{equation}
gives a big imaginary part for the scattering lengths.
At first glance the experimental situation seems to be contradictory:
The available K$^-$N-scattering
indicates a repulsive interaction while the K$^-$-atomic data 
demands an attractive potential.
The situation can be remedied by 
taking care of the existence of the $\Lambda (1405)$-resonance 
just below threshold.
Recently an improved fit of K$^-$-atomic data was carried out assuming
a nonlinear density dependence of the effective $t$-matrix \cite{Fried93}.
It has been shown that
the real part of the antikaon optical potential can be as attractive as
\begin{equation}
U^{\bar K N}_{\rm opt} \approx -200 \pm 20 \mbox{ MeV}
\end{equation}
at normal nuclear matter density while being slightly repulsive at
very low densities in accordance with K$^-$p-scattering.
The change of the sign and the nonlinear density dependance
results from the $\Lambda (1405)$-resonance.
Also another family of solutions have been
found with a moderate potential depth around $-50$~MeV.
Note that also the standard linear extrapolation gives only values of
about $-85$~MeV \cite{Fried93}. These latter two solutions
are not getting repulsive at low densities, i.e.\ fulfilling the low-density
theorem.

The K$^-$N-scattering data can be explained by
vector meson exchange models where
the $\Lambda (1405)$ is a quasi-bound
state in the t-channel \cite{Sie88,Bonn90}.
In a recent paper the coupled channel analysis
of Siegel and Weise \cite{Sie88} has been also applied for 
interactions terms coming from chiral perturbation
theory \cite{Kai95}. 
The coupled channel formalism automatically generates
the $\Lambda(1405)$ and successfully describes 
the low energy K$^-$p-scattering data.

In the following we adopt the meson-exchange picture and the chiral 
approach for the 
KN-interaction on the mean-field level and fix the parameters to
the KN-scattering length.
The case for the antikaons is then given by a G-parity transformation
which simply changes the sign of the vector potential term.
This simple treatment does not take care of the important contribution
of the $\Lambda(1405)$ resonance. 
But there exist some hints that
this resonance seems to be less important in dense matter 
(which happens when the
antikaon energy is shifted down below $m(\Lambda(1405))-m_N \approx 466$ MeV).
In ref.\ \cite{Koch94} a separable potential was applied for the
K$^-$p-interaction for finite density.
Indeed, it was found that the mass of the $\Lambda(1405)$
is shifted upwards and exceeds the K$^-$p threshold already at densities of
about $\rho\approx 0.4\rho_0$.
This is supported by recent findings within a chiral approach \cite{Waas96},
where this resonances vanishes at very low densities 
$\rho\approx 0.2\rho_0$ due to Pauli-blocking effects.
In this case the use of mean-field potentials may be justified.
Hence, we simplify our calculation 
by neglecting the contributions
coming from the $\Lambda(1405)$ in the medium and treat the problem
on the tree-level using G-parity.
Nevertheless, the results presented for the NN$\to$NN$\bar{\rm K}$K case
should be taken with some care. 
More elaborated models are needed to draw final conclusions about
the in-medium property of antikaons in the medium.

\subsection{One boson-exchange approach}

We start from the following Lagrangian \cite{Sch96}
\begin{equation}
{\cal L}_{KN} = D^*_\mu \bar K D^\mu K - m_K^2 \bar K K
- g_{\sigma K} m_K \bar K K \sigma 
- g_{\delta K} m_K \bar K \vec{\tau} K \vec{\delta} 
\end{equation}
with the covariant derivative
\begin{equation}
D_\mu = \partial_\mu +
ig_{\omega K} V_\mu + ig_{\rho K} \vec{\tau}\vec{R}_\mu
\quad .
\label{eq:kovar}
\end{equation}
For completeness we also add isospin-dependent terms which couple to
an isovector-scalar field ($\delta$) and an isovector-vector field ($R_\mu$).
Note that interaction terms of the form
\begin{equation}
{\cal L}_{KN\Lambda} = -g_{KN\Lambda} \left( \bar N \tau \gamma_5 \Lambda K +
\bar \Lambda \tau \gamma_5 N \bar K \right)
\end{equation}
do not contribute on the mean-field level as they are off-diagonal
terms. We will come to this point later in more detail.

The coupling constants to the vector mesons are chosen from the
SU(3)-relations assuming ideal mixing
\begin{equation}
2g_{\omega K} = 2g_{\rho K} = g_{\pi\pi\rho} = 6.04
\end{equation}
where $g_{\pi\pi\rho}$ is fixed by the $\rho$ decay width.
The scalar coupling constants can be fixed to the s-wave KN-scattering 
lengths \cite{Sch96}.
The isospin averaged scattering length in the tree approximation 
is given by \cite{Coh89}
\begin{equation}
\bar{a}_{KN} = \frac{1}{4} a_0^{I=0} + \frac{3}{4} a_0^{I=1} =
\frac{m_K}{4\pi \left( 1+ m_K/m_N \right)} \left(
\frac{ g_{\sigma K}g_{\sigma N} }{ m_\sigma^2 }
-2\frac{ g_{\omega K}g_{\omega N} }{ m_\omega^2 }
\right) = -0.255 \mbox{ fm}
\end{equation}
where only the isoscalar terms contribute.
This can be used to fix $g_{\sigma K}$ for known $g_{\omega K}=3.02$.
The KN-scattering lengths for a given Isospin $I$
on the tree level are then given by \cite{Coh89}
\begin{eqnarray}
a_0^{I=1} &=& \frac{m_K}{4\pi \left( 1+ m_K/m_N \right)} \left(
\frac{ g_{\sigma K}g_{\sigma N} }{ m_\sigma^2 }
+ \frac{ g_{\delta K}g_{\delta N} }{ m_\delta^2 }
-2\frac{ g_{\omega K}g_{\omega N} }{ m_\omega^2 }
-2\frac{ g_{\rho K}g_{\rho N} }{ m_\rho^2 }
\right) \label{eq:kna0i1} \\
a_0^{I=0} &=& \frac{m_K}{4\pi \left( 1+ m_K/m_N \right)} \left(
\frac{ g_{\sigma K}g_{\sigma N} }{ m_\sigma^2 }
-3\frac{ g_{\delta K}g_{\delta N} }{ m_\delta^2 }
-2\frac{ g_{\omega K}g_{\omega N} }{ m_\omega^2 }
+6\frac{ g_{\rho K}g_{\rho N} }{ m_\rho^2 }
\right) \label{eq:kna0i0}
\quad .
\end{eqnarray}
Recent experimental values are $a_0^{I=1} = 0.31$ fm and 
$a_0^{I=0} = -0.09$ fm \cite{Bar94}.
The importance of the $\delta$-meson exchange contribution can be seen
by looking at the $a_0^{I=0}$ scattering length.
The vector terms largely cancel each other as
$g_{\omega K}g_{\omega N} \approx 3 g_{\rho K}g_{\rho N}$.
Hence, without the contribution from the $\delta$-exchange
one gets
\begin{equation}
a_0^{I=0} = \frac{m_K}{4\pi \left( 1+ m_K/m_N \right)} \left(
\frac{ g_{\sigma K}g_{\sigma N} }{ m_\sigma^2 } \right) \approx 0.4 
\mbox{ fm}
\end{equation}
in contradiction with experiment
(here we used $g_{\sigma N}=10$, $g_{\omega N} = 13$ as standard values
for the RMF model).
Including the $\delta$-meson term and using
$g_{\delta N}=5.95$ from the Bonn model \cite{Mach87}
one can fit both scattering lengths nicely for
\begin{equation}
g_{\sigma K} \approx  1.9 - 2.3 \; ,  \quad  
g_{\delta K}  \approx 5.6 - 6.4
\label{eq:kncoupl}
\quad 
\end{equation}
for the various nucleonic parameter sets used in the literature
(see Table~\ref{tab1}).
Note that the values of $g_{\sigma K}$ significantly deviate
from the simple quark-model 
(simple quark counting gives $g_{\sigma K} = g_{\sigma N}/3 \approx 3.3$).
The coupling of the kaon to the $\delta$-meson is quite
strong. Therefore, we expect some effects for isospin-asymmetric 
systems which we will discuss later.

The fit based on the adjustment to the KN-scattering lengths
leads to an optical potential of
\begin{equation}
2m_K U^{\bar K}_{\rm opt} = \omega_{\bar K}^2 - m_K^2 =
g_{\sigma K}\sigma m_K 
- 2 g_{\omega K} \omega_{\bar K} V_0 - (g_{\omega K} V_0)^2
\label{eq:optpot}
\end{equation}
which gives
$U^{\bar K}_{\rm opt} = -(85\div 100)$ MeV at
normal nuclear density for the parameter sets used.
These values are lower than the ones quoted in our previous
work \cite{Sch96} as we use here the vacuum kaon mass $m_K$ instead
of the reduced mass $\mu_{KN}$. We think that this is more
consistent with the parametrisation used in the study of Kaonic atoms
\cite{Fried93}, but now our value is much closer to the standard
fit which gives $U^{\bar K}_{\rm opt} = -85$ MeV.
Note that the optical potential as defined in (\ref{eq:optpot})
is always lower than the relativistic potential the kaon feels
at normal nuclear density which is about
\begin{equation}
U^{\bar K}_{\rm rel.} = \omega_{\bar K} - m_K \approx
-(95\div 110) \mbox{ MeV} \quad .
\end{equation}
This definition corresponds to the sum of scalar and vector
potentials as discussed in \cite{Brown96b}. Nevertheless,
the scalar and also the vector potential are much lower
than the ones deduced from simple quark model counting as used 
in \cite{Brown96b}. The reason is that
our coupling ratios are about $g_{\sigma K}/g_{\sigma N} \approx 1/5$ and
$g_{\omega K}/g_{\omega N}\approx 0.23$ (see Table~\ref{tab1})
which significantly deviates from the simple quark model value of 1/3.

We have also studied the influences of off-shell terms which
have only small influences on the in-medium behaviour of kaons
(see \cite{Sch96}). Note that off-shell terms are not needed for
describing the s-wave KN-scattering lengths correctly.
On the other hand, they are essential for the chiral approach
which we will discuss in the next section.

The equation of motion for kaons in the mean-field approximation 
in uniform matter reads
$$
\left\{ \partial_\mu\partial^\mu + m_K^2
+ g_{\sigma K} m_K \sigma + g_{\delta K} m_K \tau_0 \delta_0 
+ 2(g_{\omega K} V_0 + g_{\rho K} \tau_0 R_{0,0}) 
 i\partial^\mu \right. 
$$
\begin{equation}
\left. - (g_{\omega K} V_0 + g_{\rho K} \tau_0 R_{0,0})^2
\right\} K = 0 \quad .
\label{eq:KGkaon}
\end{equation}
Note that 
there appears terms quadratic in the vector fields in eq.\ (\ref{eq:KGkaon}).
The importance of the isospin dependent terms can be estimated from
the equation of motions for the vector fields in uniform matter
\begin{eqnarray}
m_\omega^2 V_0 + d V_0^3 &=& g_{\omega N} \rho_N 
\label{eq:vectoreom} \\
m_\rho^2 R_{0,0} &=& g_{\rho N} (\rho_p-\rho_n) 
\end{eqnarray}
where $\rho_p$ and $\rho_n$ are the densities of protons and neutrons,
respectively. For led one gets  
$\rho_p-\rho_n \approx (2Z-A)/A \rho_N \approx -0.21\rho_N$,
and the isovector correction is then about
$$
\frac{g_{\rho K} R_{0,0}}{g_{\omega K} V_0} \approx 
21 \% \frac{g_{\rho N} m_\omega^2}{g_{\omega N} m_\rho^2}
\approx 8 \% 
$$
if one neglects the vector field selfinteraction which holds 
for low densities. 
Hence, the isovector contributions are
expected to be small for the densities considered here
($\rho<3\rho_0$).

The effective mass of the kaon is given by
\begin{equation}
m_K^* = \sqrt{m_K^2 + m_K \left(g_{\sigma K} \sigma + g_{\delta K}
\tau_0 \delta \right)} \quad .
\end{equation}
The scalar field $\sigma$ reduces the effective mass of the kaon in the medium,
i.e.\ the scalar interaction is attractive. The isovector-scalar field
$\delta$ shifts the effective mass if there is an isospin asymmetry in
the system.
Note that for kaons as bosons the dependence on the scalar
potential is different from that for nucleons (fermions):
for low densities
the reduction of the kaon mass in the medium is proportional 
to the square root of the scalar attraction while it is linear 
for the case of baryons (see eq.\ (\ref{eq:nucmass})). 
Moreover we point out that the scalar
potentials always follow the {\em scalar} density as demanded by
Lorentz invariance.  
The scalar density is saturating in dense matter
to ensure the existence of a saturation point of the equation of
state. As shown in \cite{Sch94b,Maru94} these nonlinear effects 
are important already at a moderate density and causes a 
saturation of the effective kaon mass with density.
We can even go further
and say that there exists a minimum effective kaon mass.
As th effective mass of the nucleon approaches zero at high density 
($m_N^*\to 0$) in the Walecka model,
the minimum scalar field is about $\sigma_{min} = -m_N/g_{\sigma N}$
and one gets for the minimum effective kaon mass
\begin{equation}
m_{K,min}^* = \sqrt{m_K^2 + m_K g_{\sigma K} \sigma_{min}}
\approx \sqrt{m_K^2 + \frac{m_K g_{\sigma K}}{m_N g_{\sigma N}}}
\approx 390 \mbox{ MeV} 
\end{equation}
for the parameters of Table~\ref{tab1}.

Decomposing the kaon field into plane waves one obtains the following 
dispersion relation for kaons (upper sign) and 
antikaons (lower sign) in uniform matter composed of nucleons only
\begin{equation}
\omega_{K,\bar K} =
\sqrt{{m^*_K}^2 + k^2 }
\;\pm\;  \left( g_{\omega K} V_0  + g_{\rho K} \tau_0 R_{0,0} \right)
\label{eq:rmfe}
\quad .
\end{equation}
Note that due to the covariant derivative coupling scheme (\ref{eq:kovar}) 
the vector term appears linearly in the kaon energy.
The vector field is repulsive (attractive) for the kaon (antikaon)
and will dominate the behaviour in very dense matter. 
For high density the kaon (antikaon) energy is then increasing
(decreasing) as $\rho^{1/3}$ because the vector field is growing 
with $\rho^{1/3}$ if one takes 
into account the vector field selfinteraction term 
(see eq.\ (\ref{eq:vectoreom})). 
Otherwise it is changing linear in density.

\subsection{Chiral Approach}

We follow the procedure outlined in \cite{Brown94}
starting from the Nonlinear Chiral Lagrangian in next-to-leading order
\begin{eqnarray}
{\cal L}^{chiral}_{KN} &=&
- \frac{3i}{8f_K^2}\left[
\bar{N}\gamma_\mu N \left(\bar{K} \tensor{\partial^\mu} K \right)
+ \bar{N}\vec{\tau}\gamma_\mu N 
  \left(\bar{K} \vec{\tau} \tensor{\partial^\mu} K \right) \right]
\nonumber \\ &&
+\frac{\Sigma_{KN}}{f_K^2}\bar{N}N\bar{K}K
+\frac{C}{f_K^2}\bar{N}\vec{\tau}N \bar{K}\vec{\tau}K
\nonumber \\ &&
+\frac{\tilde{D}}{f_K^2}\bar{N}N 
 \left(\partial_\mu\bar{K}\partial^\mu K \right)
+\frac{\tilde{D}'}{f_K^2}\bar{N}\vec{\tau}N 
 \left(\partial_\mu\bar{K}\vec{\tau}\partial^\mu K \right)
\label{chirallagr}
\end{eqnarray}
where $f_K=93$ MeV is the kaon decay constant and
$\Sigma_{KN}$ is the $KN$ sigma term.
The first two terms are the Tomozawa--Weinberg terms and are in
leading order of the chiral expansion. These are vector interactions
terms and repulsive (attractive) for kaons (antikaons). 
The other terms are in next-to-leading order. The next two terms
are scalar interactions which will decrease the effective mass of the
kaon and antikaon. The last two terms
are the so called off-shell terms which will modify the scalar
attraction. Here one encounters striking similarity with the RMF model
as the interaction is governed by scalar and vector interactions
(see \cite{Brown96b} for a discussion about this point).
In the original paper \cite{Brown94} the authors
choose $\Sigma_{KN}\approx 2m_\pi$ in accordance with the Bonn model
\cite{Bonn90}. More recently the value $\Sigma_{KN}=450\pm 30$ MeV
is favoured according to lattice gauge calculations \cite{Brown96}.
The constant $C$ can be fixed from the Gell-Mann--Okubo mass formula
to $C=33.5$ MeV. 
The corresponding scattering lengths
\begin{eqnarray}
a^{KN}_{I=1} &=& \frac{1}{4\pi f_K^2 \left(1+m_K/m_N\right)} \left[ -m_K
+\Sigma_{KN} + C + \left(\tilde{D}+\tilde{D}' \right) m_K^2 \right] \cr
a^{KN}_{I=0} &=& \frac{1}{4\pi f_K^2 \left(1+m_K/m_N\right)} \left[
+\Sigma_{KN} - 3C + \left(\tilde{D}-3\tilde{D}' \right) m_K^2 \right]
\end{eqnarray}
determine the constants $\tilde{D}$ and $\tilde{D}'$ for a given 
$\Sigma_{KN}$ via the relations
\begin{equation}
\tilde{D} \approx 0.33/m_K - \Sigma_{KN}/m_K^2 \, , \quad
\tilde{D}' \approx 0.16/m_K - C/m_K^2 \quad .
\end{equation}
Note that the off-shell terms involving the constants $\tilde{D}$ and
$\tilde{D}'$ are essential for a correct description of the scattering
lengths (see \cite{Brown94} for details).
The equation of motion in the mean-field approximation and in uniform
matter reads
$$
\left( \partial_\mu\partial^\mu + m_K^2
- \frac{\Sigma_{KN}}{f_K^2}\rho_s 
- \frac{C}{f_K^2}\tau_0 \rho_s^{iso}
\right.
$$
\begin{equation}
\left.
+ \frac{\tilde{D}}{f_K^2}\rho_s \partial_\mu\partial^\mu  
+ \frac{\tilde{D}'}{f_K^2}\tau_0 \rho_s^{iso} \partial_\mu\partial^\mu  
+ \frac{3i}{4f_K^2}\rho_N \partial_t 
+ \frac{i}{4f_K^2}\tau_0 \rho_N^{iso} \partial_t 
\right) K = 0
\label{eq:eomchiral}
\end{equation}
where $\rho_s^{iso}=\rho_{s,p}-\rho_{s,n}$ is the scalar-isovector density and 
$\rho_N^{iso}=\rho_p - \rho_n$ the vector-isovector density
which are simply the difference of the corresponding densities of
protons and neutrons. 
The mass of the kaon is shifted by
\begin{equation}
m_K^* = \sqrt{m_K^2 - \frac{\Sigma_{KN}}{f_K^2}\rho_s
- \frac{C}{f_K^2}\tau_0 \rho_s^{iso}}
\label{eq:masschiral} 
\end{equation}
and the same arguments as for the case of the one-boson exchange model
holds. One sees again that the scalar potential for the kaon behaves
differently as the one for nucleons (\ref{eq:nucmass}). 
More important is, that there exists a minimum effective kaon mass
as a minimum scalar field implies a maximum scalar density for
the Walecka model which is about $\rho_{s,max} \approx 2\rho_0$
(see e.g.\ \cite{Rei88}). This gives a minimum kaon effective mass
of $350-400$ MeV depending on the kaon-nucleon sigma term.
The influence of the isovector terms can be estimated
from isospin considerations:
e.g.\ for led one has 
$\rho_p-\rho_n \approx (2Z-A)/A \rho_N \approx -0.21\rho_N$,
i.e.\ about $21/3 = 7\%$ correction for the vector-isovector term
of eq.\ (\ref{eq:eomchiral}). This is in accordance with our estimate
for the one-boson exchange model in the previous section.
In the following we will neglect the isovector contributions.
Fourier transformation of the equation of motion yields
\begin{equation}
-\omega^2 + k^2 + \Pi (\omega,k;\rho_N) = 
-\omega^2 + k^2 + m_K^2 - \frac{\Sigma_{KN}}{f_K^2}\rho_s
- \frac{\tilde{D}}{f_K^2}\rho_s \omega^2
- \frac{3}{4f_K^2} \omega \rho_N = 0 \quad .
\end{equation}
where $\Pi (\omega,k;\rho_N)$ is kaon self energy which depends
in general also on the kaon energy. This has to be taken into account
to get the energy of a kaon/antikaon in the nuclear medium
\begin{equation}
\omega_{K,\bar K} = \left[\sqrt{ {m_K^*}^2
\left(1+\frac{\tilde{D}}{f_K^2}\rho_s\right)
+ k^2 + \left(\frac{3}{8f_K^2} \rho_N\right)^2}
\;\pm\; \frac{3}{8f_K^2} \rho_N \right]
\left(1+\frac{\tilde{D}}{f_K^2}\rho_s\right)^{-1} \quad ,
\label{eq:chpte}
\end{equation}
where $m_K^*$ is defined by eq.\ (\ref{eq:masschiral}). 
Here we note that in the high density limit the kaon energy is growing
linear with density while for the antikaon the energy saturates
at $m^*_K$ as the vector contributions cancel each other.
The optical potential at normal nuclear density
\begin{equation}
U^{\bar K}_{\rm opt} = \frac{1}{2m_K} \Pi (\omega_{\bar K},k=0;\rho_0) 
\approx -68 \mbox{ MeV}
\end{equation}
is rather moderate while the relativistic potential is about
$-75$ MeV. This is in contrast to the findings of Brown and Rho
\cite{Brown96b} who gets a rather deep potential of $-200$ MeV.
There are several reasons for this discrepancy: first BR--scaling
is not taken into account here (which gives an additional factor of
5/3 at $\rho_0$, i.e.\ a potential of  $-125$ MeV), 
second Brown and Rho neglect the off-shell term
and do not take into account the KN-scattering lengths, third
they neglect the energy dependence of the kaon self energy, fourth
they assume that the scalar and vector density are equal, fifth
they neglect that the scalar potential of the kaon behaves differently
in matter compared to the nucleon one (see the discussion of 
eq.\ (\ref{eq:masschiral}) above).

\section{Results}

\subsection{Kaon energy in matter}

Recently, the dynamics of the $\Lambda(1405)$ has been studied in 
nuclear matter using a coupled channel formalism 
\cite{Koch94,Waas96}. The most important finding is that the
effects coming from the $\Lambda(1405)$ vanishes at rather low
densities ($\rho<0.25\rho_0$). The optical potential for the antikaon
is about $-100$ MeV \cite{Koch94} and $-107$ MeV \cite{Waas96}
corresponding to a kaon energy of $\omega_{\bar K} = 380$ MeV
and $\omega_{\bar K} = 372$ MeV at normal nuclear density,
respectively. These values are in accordance with the ones
calculated in the mean field approximation 
in the previous sections within the relativistic mean field (RMF) model and
the chiral perturbation theory (ChPT).

\begin{figure}
\epsfbox{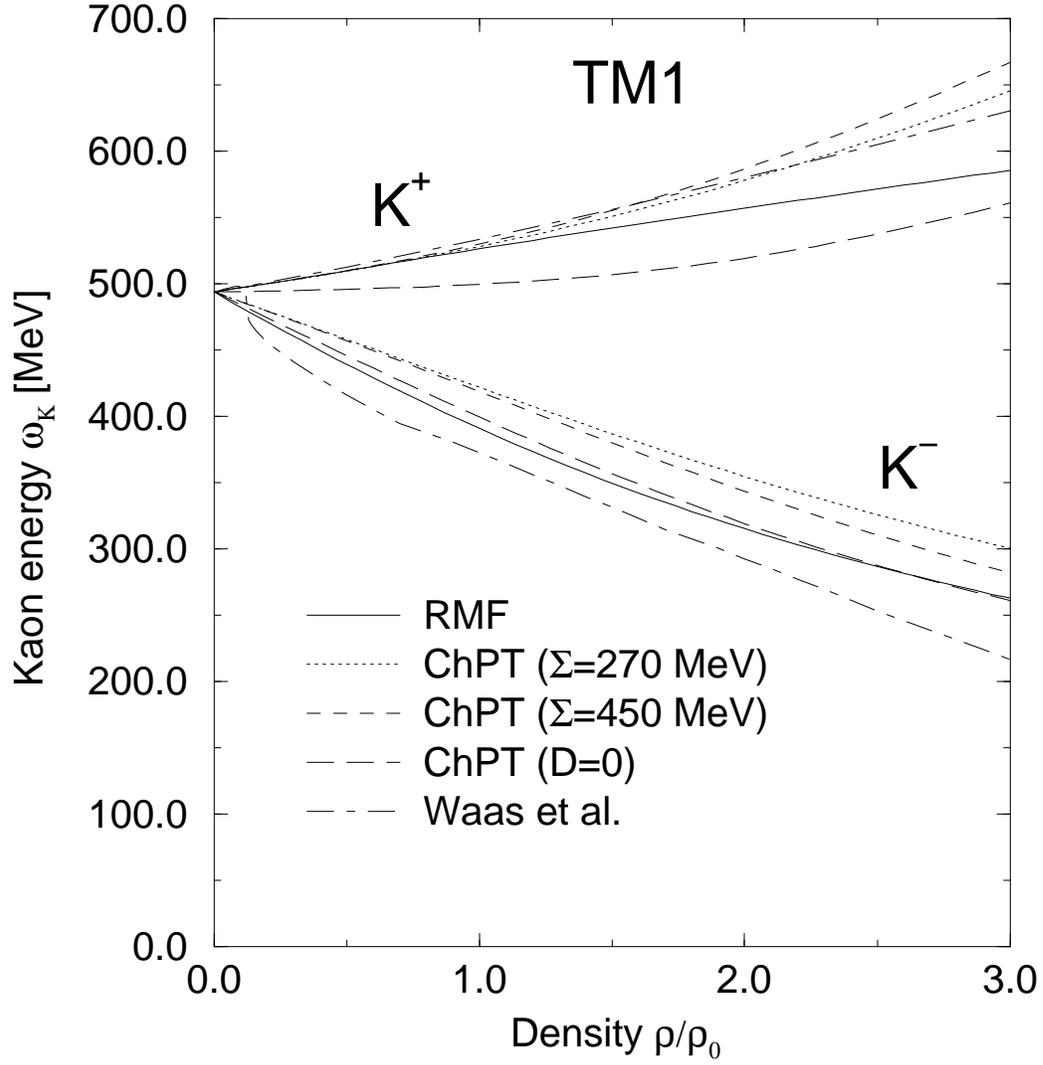}
\caption{
The energy of kaons and antikaons in nuclear matter as function of
density for the soft EOS (parameter set TM1).}
\label{fig2}
\end{figure}

In the following we discuss the in-medium energy of kaons and
antikaons in nuclear matter using a soft (set TM1 of Table~\ref{tab1})
and a hard (set NL-Z of Table~\ref{tab1}) equation of state.
Figure~\ref{fig2}
shows the energy of kaons (upper curves) and antikaons
(lower curves) with the soft EOS for the RMF model (eq.\ (\ref{eq:rmfe})),
ChPT (eq.\ (\ref{eq:chpte})) and the results of the coupled channel
analysis of Waas et al.\ \cite{Waas96}. 
In the case of ChPT we discuss three cases: i) for a sigma term of
$\Sigma_{KN} = 2m_\pi$ as used in \cite{Brown94},
ii) for a sigma term of $\Sigma_{KN} = 450$ MeV
as derived from recent lattice data \cite{Brown96},
iii) for vanishing off-shell terms (denoted as $\tilde{D} = 0$)  
and a sigma term
of $\Sigma_{KN} = 2m_\pi$ as used as input for the RBUU calculations
\cite{Li95,Fang94,kaonflow,Li94}.

All models show a quite similar behaviour in Fig.~\ref{fig1} 
for the kaon energy
at low density except for the case $\tilde{D}=0$.
This results from the low density theorem and is a generic feature
when the coupling constants are fixed to the KN-scattering lengths.
Neglecting the off-shell terms, i.e. setting $\tilde{D}=0$,
violates the low density theorem.  This gives a slower raise of the kaon
energy with density and the kaon energy nearly stays constant for a
wide range of density.
Note that this latter parametrisation for the
kaon energy is used in the dynamical calculations \cite{Fang94}. 
For higher density the other curves also start to deviate. The ChPT gives
a higher kaon mass, i.e.\ more repulsion than in the RMF model.
The results of the coupled channel calculation seems to follow 
more closely the one of ChPT. At $\rho=3\rho_0$ the kaon energy 
reads $\omega_K = 585$ MeV for the RMF model, $\omega_K=630$ MeV for
the coupled channel analysis \cite{Waas96} and
$\omega_K = 640-670$ MeV for ChPT, so they 
deviate about 85 MeV from each other. 

The antikaon energy (lower curves) of the different models
is always attractive, except for the small density region for
the coupled channel calculations due to the $\Lambda(1405)$ resonance.
The latter one gives the most attraction of about 
$\omega_{\bar K}=217$ MeV at $\rho=3\rho_0$, 
followed by the RMF model with $\omega_{\bar K}= 263$ MeV and the ChPT with
$\omega_{\bar K}=280\div300$ MeV. 
The curve for the case of $\tilde{D}=0$ used in \cite{Li94} 
follows closely the one for the RMF model.
All the curves for the antikaon energy are lying surprisingly close
together.
Note that the prediction of ChPT is rather insensitive to the choice
of $\Sigma_{KN}$ but rather sensitive to the off-shell terms,
especially for the kaon energy.

\begin{figure}
\epsfbox{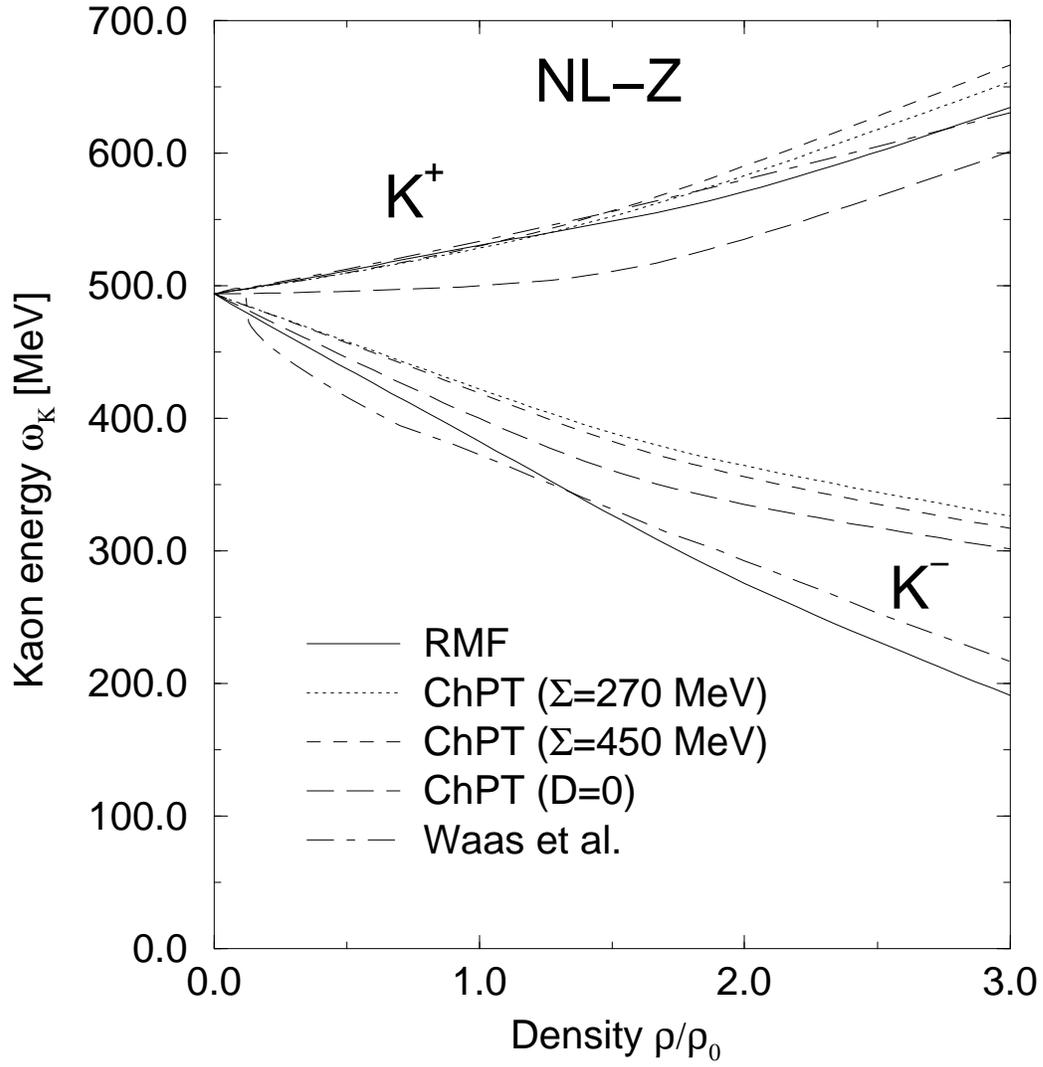}
\caption{
The same as Fig.~\protect\ref{fig2} for the hard EOS (parameter set NL-Z).
}
\label{fig3}
\end{figure}

In Fig.~\ref{fig3} the case for the hard EOS is plotted. Now the curves
of the kaon energy are lying very close together, even at higher
densities. This is due to the fact that the vector potential in the
RMF model is now raising linear with density as in the ChPT 
in contrary to the soft EOS where it raises like $\rho^{1/3}$ due to the vector
self-interaction terms. The energy of the kaon is now between
$\omega_K= 630\div 670$ MeV at $\rho=3\rho_0$. 
Without the off-shell terms,
the kaon energy significantly deviates from the other curves
and stays rather constant up to $1.5\rho_0$. Note that the overall
changes for the hard EOS compared to the soft EOS are quite moderate,
especially when using ChPT, and only show up at higher density.

The different predictions for the antikaon energy seems to split now
into two regimes: the results for the ChPT give a antikaon energy
of about $\omega_{\bar K}= 300$ MeV at $\rho=3\rho_0$ rather 
independent of the off-shell term and the choice of $\Sigma_{KN}$,
while the RMF model and the coupled channel analysis get around
$\omega_{\bar K}= 200$ MeV at $\rho=3\rho_0$.
The antikaon energy within the RMF model is now much deeper
due to the stronger vector potential compared to the soft EOS.
We want to point out again, that Dirac-Br\"uckner calculations
seems to favour the soft EOS \cite{Bod91,Gmu91}. Nevertheless,
we see that the differences in the kaon/antikaon energy 
due to the EOS are well within the
differences of the model predictions.

\subsection{Threshold energy for kaon production in matter}

In the following we discuss the shift of the threshold energy
of various processes for heavy ion collisions due to medium modifications.

\begin{figure}
\epsfbox{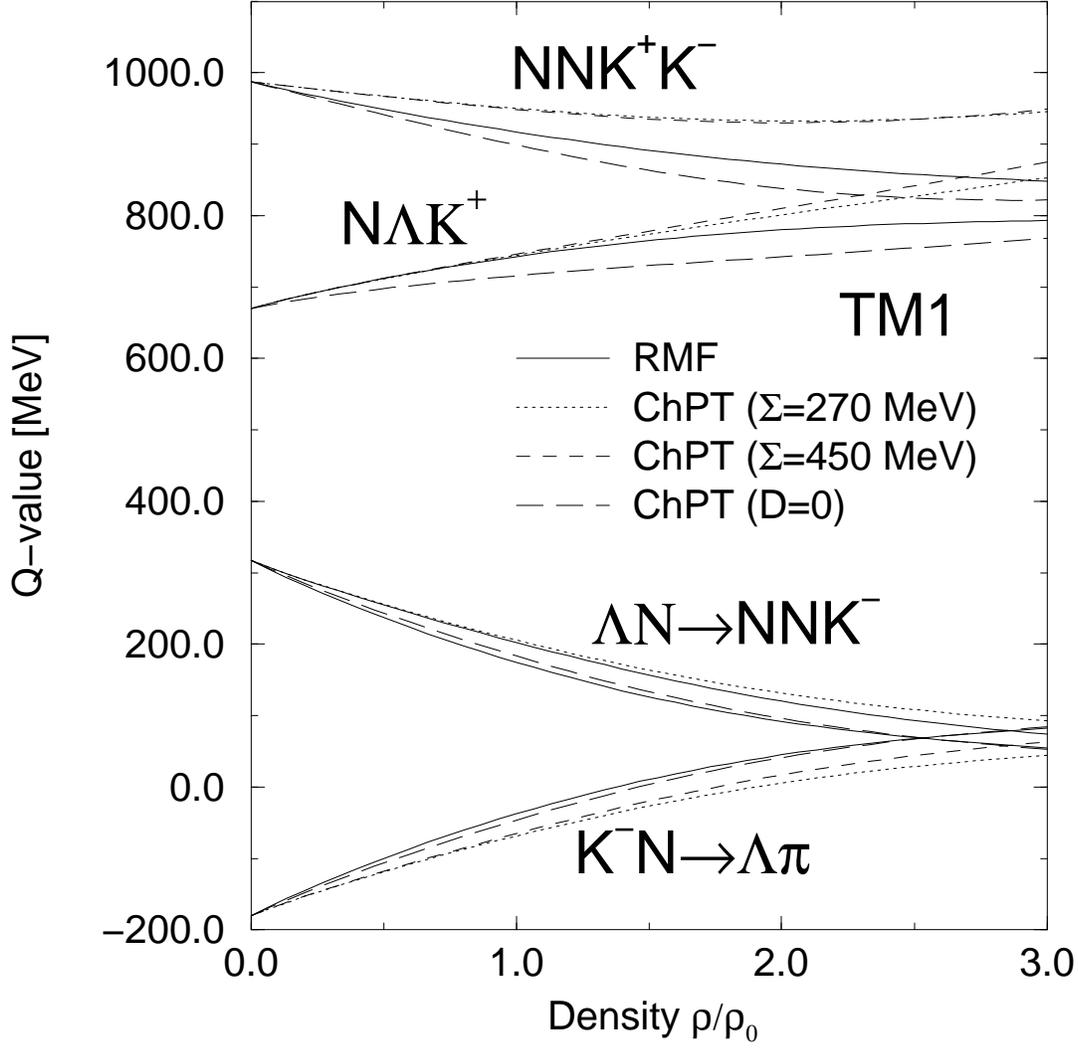}
\caption{
The $Q$-values of the production processes of kaons and antikaons
(the two upper bunches of curves) and the annihilation processes
of antikaons and $\Lambda$'s (lower bunches of curves)
versus the density for the soft EOS.}
\label{fig4}
\end{figure}

\begin{figure}
\epsfbox{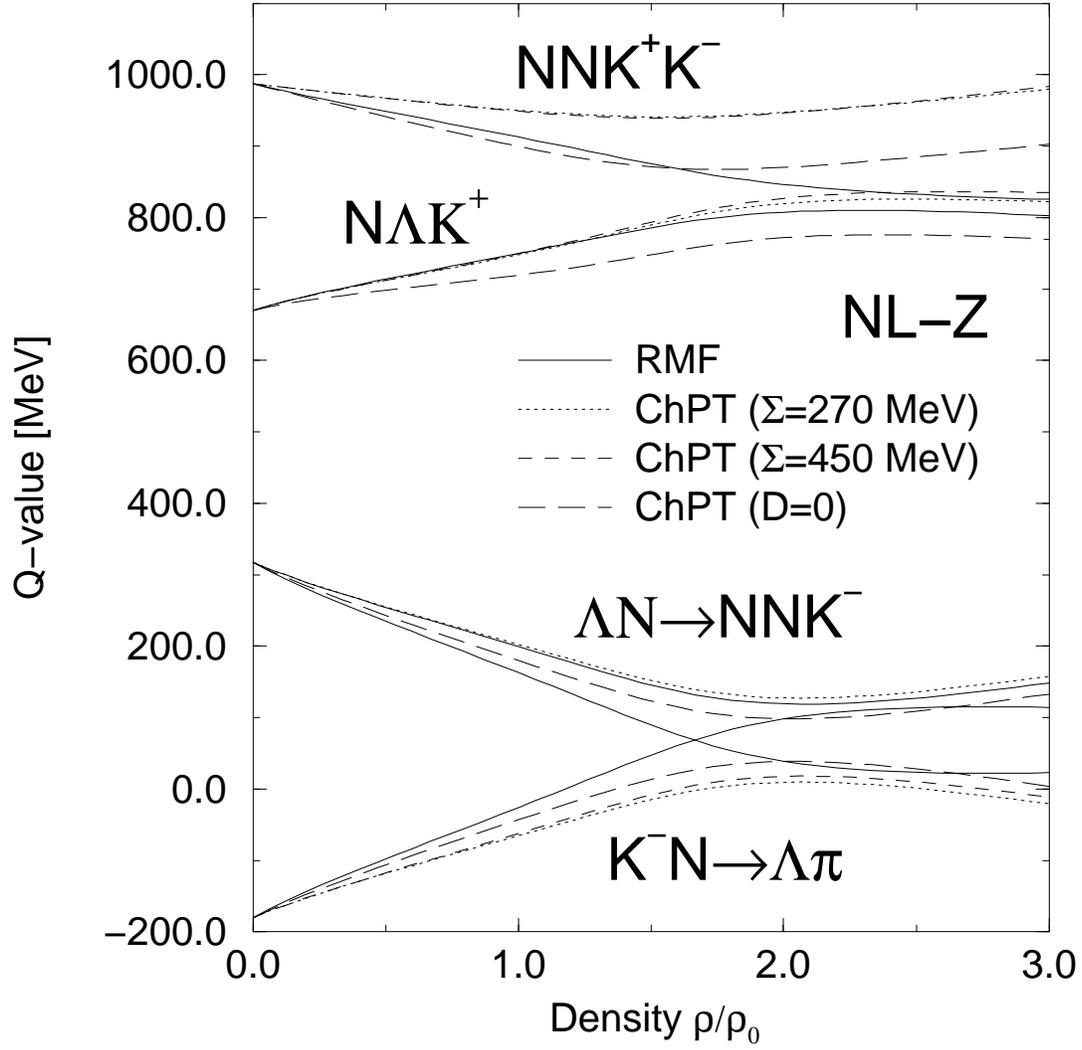}
\caption{
The same as Fig.~\protect\ref{fig4} for the hard EOS.
}
\label{fig5}
\end{figure}

Kaons are mainly produced at threshold via the process
NN$\to$N$\Lambda$K. The minimum energy needed is 
$Q({\rm N}\Lambda{\rm K})\approx 671$ MeV in vacuum.
In the medium, the threshold is shifted to
\begin{equation}
Q({\rm N}\Lambda{\rm K})=E_\Lambda(p=0) + \omega_K(k=0) - E_N(p=0)
\end{equation}
where we assume that the outgoing nucleon is not Pauli-blocked in 
the hot zone of the collision. Hence, the subthreshold production
of kaons is sensitive to three different in-medium effects:
the EOS ($E_N$), the $\Lambda$ potential ($E_\Lambda$) and the
kaon energy ($\omega_K$) in medium. 
These effects will partly cancel each other as the kaon feels a
repulsive potential of $29$ MeV (eq.\ (\ref{eq:kaonpot})) while
the $\Lambda$ sees an attractive potential of $-30$ MeV at $\rho_0$
(eq.\ (\ref{eq:lambdapot})). Therefore, subthreshold kaon production 
seems to probe mainly the EOS. As the nucleons feel an attractive
potential of about $-60$ MeV the threshold will be shifted {\em
upwards} at normal nuclear density by this amount and the production
of kaons is {\em reduced} in the medium. This is indeed the case
as can be seen from Fig.~\ref{fig4} which shows the
threshold energy $Q({\rm N}\Lambda{\rm K})$ as a function of density.
The similar behaviour of the different curves at low density is due
to the low-density theorem. 
At $\rho=3\rho_0$ the value of $Q({\rm N}\Lambda{\rm K})$ reaches
about 800 MeV for the RMF model and about 860 MeV for ChPT which is
quite insensitive to the value of the sigma term.
Without off-shell terms, the threshold
energy is underestimated in medium, and we expect that the production
rates for kaons calculated in \cite{Fang94,Li94} are overestimated.
Note, that all calculations ignoring in-medium effects
\cite{Hart94,Maru94b} will also
give a too high production rate for kaons.
The case for the hard EOS is plotted in Fig.~\ref{fig5}.
The behaviour of the threshold energy in medium is quite similar
for the different EOS considered here. 
Again, the low density limit more or less fixes the shape of the
curves of the kaon energy independent of the EOS used. 
The curves for the RMF model and ChPT are lying closely 
between $800-830$ MeV at $\rho=3\rho_0$. Especially, the curve
for the RMF model does not change considerable for the hard EOS
compared to the soft one.
All curves seem to saturate
for the hard EOS but are lying within the uncertainties of the
different models used for the kaon energy. 
A definite conclusion
whether or not subthreshold kaon production probes the EOS can not be
drawn until the in-medium properties of the kaon can be determined
more precisely.

Antikaons are created in heavy ion collisions first by
the process NN$\to$NNK$\bar K$. The threshold value of 
$Q({\rm K}\bar K)\approx 988$ MeV is modified in the medium by
the sum of the kaon and antikaon energy 
$Q({\rm K}\bar K)=\omega_K(k=0) + \omega_{\bar K}(k=0)$.
Therefore, subthreshold antikaon production probes the in-medium
property of kaons and antikaons solely. As the vector potential
cancels out approximately, it will mainly depend on the scalar potential
the kaon feels in the medium. The upper curves in Fig.~\ref{fig4}
show that indeed 
$Q({\rm K}\bar K)$ is reduced in the medium in all models discussed
here. ChPT predicts an in-medium reduction of about $-56$ MeV at
maximum compared to the vacuum and then the curves go up again
for higher density. The reason is that the sum of the kaon and antikaon
energy contains a term coming form the Tomozawa--Weinberg term
\begin{equation}
\omega_K(k=0) + \omega_{\bar K}(k=0) = 2\sqrt{ {m_K^*}^2
\left(1+\frac{\tilde{D}}{f_K^2}\rho_s\right)
+ \left(\frac{3}{8f_K^2} \rho_N\right)^2}
\left(1+\frac{\tilde{D}}{f_K^2}\rho_s\right)^{-1} \quad ,
\end{equation}
which is repulsive and dominates at higher density.
On the other side, the RMF model gives a reduction of about
$-140$ MeV at $\rho=3\rho_0$. The $Q$-value is steadily decreasing
as the sum of the kaon and antikaon energy
\begin{equation}
\omega_K(k=0) + \omega_{\bar K}(k=0) =
2m^*_K = 2\sqrt{m_K^2 + m_K g_{\sigma K} \sigma}
\end{equation}
depends on the attractive scalar potential only.
The curve used in RBUU calculations with a soft EOS 
\cite{Li94} is lying even lower
and hence, the production rates of antikaons seems to be
overestimated. 
Using the hard EOS (Fig.~\ref{fig5}) the situation does not change
significantly. The $Q$-value in the RMF model is now reduced by $-160$ MeV
at $\rho=3\rho_0$. The curves for the ChPT go up stronger at
higher density compared to the soft EOS 
as they are sensitive to the strength of the vector potential (i.e.\ to
the behaviour of the EOS at high density)
in contrast to the RMF model.

As an interesting fact, the $Q$-values for
kaon and antikaon production are lying close together for the RMF
model. Note that this does not mean that the numbers of produced
kaons and antikaons are the same inside the dense medium. 
The kaons will be produced at different density and the average
$Q$-value over the density profile will give a measure for the produced
kaons and antikaons in the medium. On the other hand,
the production of kaons will be dominated by the secondary
processes (rescattering effects) N$\Delta\to$N$\Lambda$K and 
$\pi$N$\to\Lambda$K, the production of antikaons by
the processes N$\Delta\to$NNK$\bar K$ and  $\pi$N$\to$K$\bar K$.
Let us assume that the change of the $\Delta$ mass and energy 
is equal to that of the nucleon. Then the $Q$-values of 
these channels can be simply derived by shifting the 
corresponding curves for the $Q$-values of Figs.~\ref{fig4} and \ref{fig5}
down by $m_N-m_\Delta\approx -290$ MeV (ignoring the finite width of the
$\Delta$) and by $-m_\pi$, respectively. If the $\Delta$ feels a higher
(lower) potential than the nucleon, then this will suppress (enhance)
subthreshold kaon production. Processes involving two $\Delta$'s in
the entrance channel will decrease the $Q$-value by $-580$ MeV compared to the
two nucleon one and hence, enhanced production of kaons  
will be sensitive to $\Delta$ matter (density isomers) \cite{Hart94}.

Also annihilation processes will play a dominant role at high density.
Kaons will not annihilate and escape due to their long free mean path.
But the charge exchange reaction ${\rm K^+ n\to K^0 p}$ 
will act like an annihilation process for kaons
as only charged particles are measured in the present heavy ion experiments
\cite{GSI}. This process will be modified in the medium only by
isovector potentials. We do not expect changes of the threshold energy
for isospin symmetric systems. As the isospin potential for the kaon is
negligible (see eq.\ \ref{eq:isoopt}) the threshold energy will be only 
shifted by the isovector potential of the nucleons. 
The maximum effect will be seen for systems like led where one gets
\begin{equation}
Q({\rm iso})\approx E_n(p=0)-E_p(p=0) = -2g_{\rho N} R_{0,0} 
= -2\frac{g_{\rho N}^2}{m_\rho^2} \rho_{\rm iso} \approx 
(16\div 19) \mbox{ MeV} \frac{\rho_N}{\rho_0}
\end{equation}
with the parameters of Table~\ref{tab1}. Hence, the charge exchange
process will be a little bit suppressed in isospin asymmetric systems.
The change is quite moderate but comparable with 
the in-medium shift of the $Q$-value
for the kaon production process. 

Antikaons will annihilate strongly due to the process 
$\bar K$N$\to\Lambda\pi$ which is exothermal in vacuum 
($Q(\bar K$N$\to\Lambda\pi)\approx -180$ MeV). Also the $\Lambda$ can
annihilate via the process $\Lambda$n$\to$NN$\bar K$, but this process
is endothermal ($Q(\Lambda$N$\to$NN$\bar K)\approx 317$ MeV) in vacuum
and is usually neglected. Nevertheless, we expect rather strong
in-medium modifications of these $Q$-values as the antikaon energy is
involved which changes considerably in nuclear matter.
The lower curves in Fig.~\ref{fig4} and \ref{fig5} show these
$Q$-values as a function of density. All models give an astonishingly
similar strong behaviour in dense matter: 
$Q(\bar K$N$\to\Lambda\pi)$ is going up with
density and crosses zero at $\rho\approx1.5\rho_0$ while
$Q(\Lambda$N$\to$NN$\bar K)$ is decreasing rapidly. For the soft EOS,
both $Q$-values reaches even similar values at high density of
about 70 MeV regardless of the model used.
This means that the annihilation of $\Lambda$'s is favoured in the
medium while the annihilation of antikaons is suppressed.
At very high density these processes are even equally possible.
For the hard EOS (Fig.~\ref{fig5}) the $Q$-values for the ChPT
seem to saturate at high density at 
$Q(\bar K$N$\to\Lambda\pi)\approx 0$ MeV and
$Q(\Lambda$N$\to$NN$\bar K)\approx 140$ MeV. On the contrary, the
curves for the RMF model show a crossing, so that the situation is reversed
and one gets $Q(\bar K$N$\to\Lambda\pi)\approx 115$ MeV
and $Q(\Lambda$N$\to$NN$\bar K)\approx 20$ MeV.
It would be interesting to examine how these in-medium effects of
the annihilation process will influence the antikaon and $\Lambda$
spectra in heavy ion collisions at subthreshold energy
where they are most pronounced.

\section{Conclusions and Outlook}

The study of the in-medium properties of kaons and antikaons shows
that it is important to link the models to the available data, here to
the KN scattering data, and to 
take into account effects nonlinear in density. Then one gets 
rather similar predictions in the models discussed here for the 
energy of kaons and antikaons in nuclear matter up to a certain
density, say $(1-2)\rho_0$. The kaon energy in matter 
is well determined by the low density theorem, 
while the antikaon energy is more model dependent.
The threshold energy for the production of kaons is shifted up
in dense matter, while the one for antikaons is considerable
decreased. Also the threshold energy for the 
annihilation processes for antikaons and
$\Lambda$'s show strong in-medium modification and can even get
similar values at high density. Hence, it will be important to study
the process $\Lambda$N$\to$NN$\bar K$ in the medium which will enhance
the number of produced antikaons in heavy ion collisions at
subthreshold energy. 
This will also change the flow pattern of antikaons and $\Lambda$'s
and will cause e.g.\ an antiflow of $\Lambda$'s for central rapidities.
As the threshold energy for antikaon and $\Lambda$ production as well
as for annihilation become equal around $\rho\approx 3\rho_0$ 
the number of antikaons and $\Lambda$'s will be predicted to be
equal in the dense zone of a heavy ion collision at subthreshold
energy due to in-medium effects. The number of kaons will then be
twice the number of antikaons due to strangeness conservation.
This effect might be seen at midrapidity and high momenta.

Insofar, we have only discussed effects on the mean-field level
which will cause shifts of the threshold energy and essentially modify
the phase space of the reactions in the medium. Using Fermis golden
rule, one can now implement these modifications into a dynamical model
by simply changing the energy of all hadrons consistently and leaving
the cross sections constant.
Nevertheless, also the cross sections might change in the medium.
For the process NN$\to$N$\Lambda$K, the N$\Lambda$K vertex has to be
considered which vanishes on the mean field level. Hence, changes of
the cross section are here of higher order. They have to be computed by
taking into account the p-wave interactions of nucleons and kaons
and will change the angular distribution of the produced kaons
in heavy ion reactions.
The investigation of these effects is an interesting task and 
will be considered in a forthcoming work.

\acknowledgements

We like to thank Avraham Gal for initiating the study of the
nonlinear behaviour of the $\Lambda$ potential with density.
We also thank Bo Jakobsson
and Wolfram Weise for useful discussions and remarks.
J.S. thanks Thomas Waas for the submission of the results
of the coupled channel calculation for the kaon/antikaon energy in medium and
J\"org Aichelin, Eckart Grosse, Christoph Hartnack, 
Che Ming Ko, Peter Senger and Andreas Wagner for illuminating conversations.
I.N.M. is supported in part by the International Science Foundation (Soros)
grant N8Z000 and EU-INTAS grant 94-3405.
Two of us (J.S. and I.N.M.) expresses their gratitude to the Niels Bohr
Institute for their warm hospitality and financial support.

\begin{table}

\caption{The coupling constants of the parameter sets used.
The vector coupling constant for the $\Lambda$ are taken from
SU(6)-relations. 
The coupling constants of the kaons to the 
$\sigma$- and $\delta$-meson are fixed
by the s-wave KN-scattering lengths. The vector coupling constants
are chosen from SU(3)-relations.
The parameters for the scalar and vector selfinteraction
terms are not given, they can be found in the corresponding references.}
\begin{center}
\begin{tabular}{ccccccc}
Set & NL-Z  &  NL-SH &    PL-Z  &    PL-40 &    TM1   &   TM2   \cr
\hline
Ref. & \cite{Rufa88} & \cite{Shar93} & \cite{Rei88} & \cite{Rei88} &
\cite{Toki94} & \cite{Toki94} \cr
\hline
$g_{\sigma N}$ &          10.0553 &  10.4440 &  10.4262 &  10.0514 &  10.0289 &
11.4694 \cr
$g_{\omega N}$ &          12.9086 &  12.9450 &  13.3415 &  12.8861 &  12.6139 &
14.6377 \cr
$g_{\rho N}$ &             4.8494 &   4.3830 &   4.5592 &   4.8101 &   4.6322 &
4.6783 \cr
\hline
$g_{\sigma \Lambda}$ &     6.23 &     6.47 &     6.41 &     6.20 &     6.21 &
7.15 \cr
$g_{\omega \Lambda}$ &     8.61 &     8.63 &     8.89 &     8.59 &     8.41 &
9.76 \cr
\hline
$g_{\sigma K}$   &  1.85 &  2.05 &   2.20 &   2.27 &   1.93 &   2.27 \cr
$g_{\omega K}$   &  3.02 &  3.02 &   3.02 &   3.02 &   3.02 &   3.02 \cr
$g_{\rho K}$     &  3.02 &  3.02 &   3.02 &   3.02 &   3.02 &   3.02 \cr
$g_{\delta K}$   &  6.37 &  5.59 &   5.89 &   6.31 &   5.87 &   5.94 
\end{tabular}
\end{center}
\label{tab1}
\end{table}

\end{document}